\documentclass[preprint, showpacs, preprintnumbers, superscriptaddress]{revtex4}

\usepackage{graphicx}
\usepackage{dcolumn}
\usepackage{bm}
\usepackage[T1]{fontenc} 
\newcommand{\SMO }{SrMnO$_3$}
\newcommand{\LSMO }{La$_{0.7}$Sr$_{0.3}$MnO$_3$}
\newcommand{\LS }{La$_{0.7}$Sr$_{0.3}$O}

\newcommand{\STO }{SrTiO$_3$}
\newcommand{\NSTO}{Nb:SrTiO$_3$}
\newcommand{\IV}{\textit{I}-\textit{V}}
\newcommand{\CV}{\textit{C}-\textit{V}}
\begin{document}

\preprint{APS/123-QED}

\title{Termination Control of the Interface Dipole in La$_{0.7}$Sr$_{0.3}$MnO$_3$/Nb:SrTiO$_3$ (001) Schottky Junctions}

\author{Yasuyuki Hikita}
 \email{hikita@k.u-tokyo.ac.jp}
 \affiliation{Department of Advanced Materials Science, University of Tokyo, Kashiwa, Chiba 277-8561, Japan}
\author{Mitsuru Nishikawa}
 \affiliation{Department of Applied Physics, University of Tokyo, Tokyo 113-8656, Japan}
\author{Takeaki Yajima}%
 \affiliation{Department of Advanced Materials Science, University of Tokyo, Kashiwa, Chiba 277-8561, Japan}
\author{Harold Y. Hwang}%
  \affiliation{Department of Advanced Materials Science, University of Tokyo, Kashiwa, Chiba 277-8561, Japan}
  \affiliation{Department of Applied Physics, University of Tokyo, Tokyo 113-8656, Japan}
  \affiliation{Japan Science and Technology Agency, Kawaguchi, Saitama 332-0012, Japan}

\date{\today}

\begin{abstract}
In order to investigate the interface termination dependence of perovskite band alignments, we have studied the Schottky barrier height at La$_{0.7}$Sr$_{0.3}$MnO$_3$/Nb:SrTiO$_3$ (001) heterointerfaces. As the Nb:SrTiO$_3$ semiconductor was varied from TiO$_2$ termination to SrO termination by variable insertion of a SrMnO$_3$ layer, a large systematic increase in the Schottky barrier height was observed. This can be ascribed to the evolution of the interface dipole induced to screen the polar discontinuity at the interface, which gives a large internal degree of freedom for tuning band diagrams in oxides.
\end{abstract}

\pacs{73.40.Sx, 73.40.Ei, 73.40.Cg}
\maketitle
There has been burgeoning recent interest in the electronic structure of complex oxide heterointerfaces. Technical advances in oxide thin film growth allow the fabrication of structures with atomic scale precision, and together with theoretical advances, a host of new interface electronic states have been found and/or predicted ~\cite{SOkamoto1, Ohtomo, ZSPopvic, WCLee, RPentcheva, NReynen, ARuegg, Smadici, JChaloupka}. Fundamental to this endeavor is knowledge of band lineups, which is crucial for the design of new interface states ~\cite{WCLee}, as well as the engineering of oxide devices. Compared to conventional semiconductors and metals, however, oxide heterointerfaces are far less understood ~\cite{Okamoto}.
\\\indent
\begin{figure}[t]
  \begin{center}
    \includegraphics[clip]{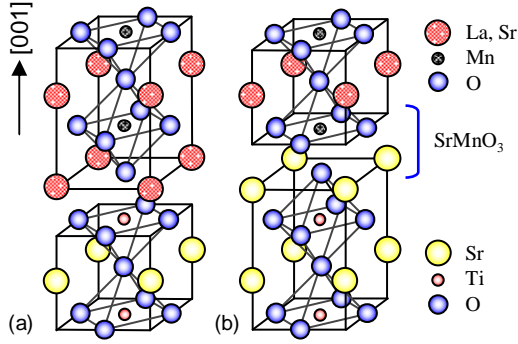}
  \end{center}
  \caption{(Color online) The two different possible interfaces between the perovskites \LSMO\ and \STO\ joined in the [001] direction: (a) TiO$_2$-terminated, and (b) SrO-terminated \STO.}
  \label{fig:Fig1.eps}
\end{figure}
In addition to improving the basic knowledge of complex oxide work functions, electron affinities, etc., there are also structural degrees of freedom at their interfaces which have been little explored. For example, the heterointerface between two (001)-oriented perovskites with different cations can have two different interface terminations (Fig. \ref{fig:Fig1.eps}). Given the partially ionic nature of oxides, the different terminations could have significantly different interface dipoles, thus changing the band lineup across the interface. In order to experimentally investigate this effect, we have studied the Schottky interface between the ferromagnetic metal \LSMO\ and the \textit{n}-type semiconductor Nb-doped \STO\ (0.05 wt \% doped). In addition to providing a model system for these studies, this interface is of strong interest in magnetic tunnel junctions ~\cite{JZSun}, magnetic field sensitive diodes ~\cite{Nakagawa}, and for enhanced photocarrier injection ~\cite{Katsu}.
\\\indent
In this paper, we report the investigation of the Schottky barrier height (SBH) in (001)-oriented \LSMO/\NSTO\ Schottky junctions as 0 - 1 unit cell (uc) of \SMO\ is inserted at the interface. By growing a \LSMO\ film directly on TiO$_2$-terminated \NSTO, a MnO$_2$/\textit{La$_{0.7}$Sr$_{0.3}$O}/TiO$_2$ interface is formed [Fig. \ref{fig:Fig1.eps}(a)]. Alternatively, by first growing 1 uc of \SMO\ before \LSMO\ deposition, the MnO$_2$/\textit{SrO}/TiO$_2$ interface is formed [Fig. \ref{fig:Fig1.eps}(b)], which is equivalent to the deposition of \LSMO\ on the alternative SrO-terminated \NSTO\ surface ~\cite{Hotta}. The deposition of a fractional unit cell of \SMO\ allows the study of the evolution of the SBH between these endpoints, which was probed using current-voltage (\textit{I-V}), capacitance-voltage (\textit{C-V}), and internal photoemission (IPE) measurements. All experiments indicate a systematic increase in the SBH by changing the termination layer at the interface. Although this result is difficult to understand within a Schottky-Mott ~\cite{Sze} or Bardeen ~\cite{Bardeen} framework for metal-semiconductor interfaces, a simple consideration of the evolution of the screening dipole at the interface explains this trend, which is expected to be quite general for metal-semiconductor and metal-insulator perovskite heterointerfaces.
\\\indent
The heterojunctions were fabricated by pulsed laser deposition using a KrF excimer laser with a laser fluence of 0.22 J/cm$^2$, substrate temperature of 850 $^{\circ}$C, and an oxygen partial pressure of 1 $\times$ 10$^{-3}$ Torr, as previously optimized ~\cite{Song} . The interface termination was varied by deposition of a calculated thickness of \SMO\ (\SMO \ = 0.0, 0.3, 0.6, 1.0 uc), for which the deposition rate was calibrated with reflection high energy electron diffraction (RHEED) prior to the fabrication of the final structures. After the deposition of \SMO, 100 uc of \LSMO\ was deposited by monitoring the RHEED oscillations.  Temperature dependent magnetization measurements give Curie temperatures $T_C$ $\sim$ 360 K in all cases. Ohmic contacts to the \LSMO\ and the \NSTO\ were made by evaporated gold films and In ultrasonic soldering, respectively, contacting an array of junctions each $\sim$ 0.25 mm$^2$ in area. All the measurements were carried out at room temperature and the polarity of the applied bias is defined as positive when applied to the \LSMO.
\\\indent
\begin{figure}[t]
  \begin{center}
    \includegraphics[clip]{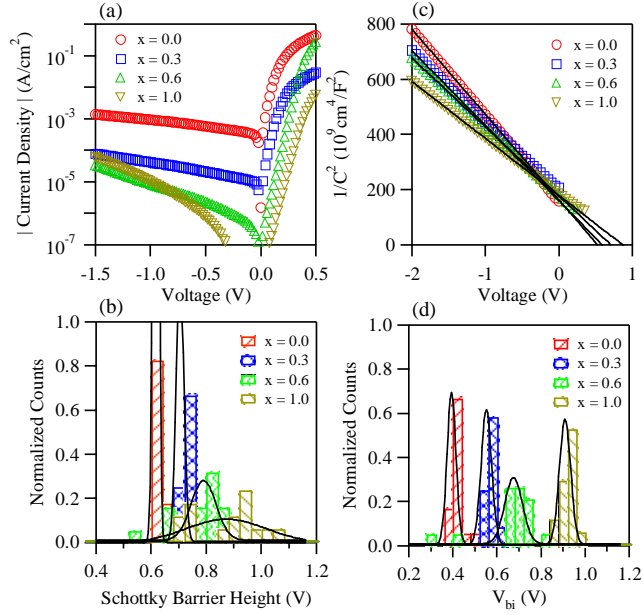}
  \end{center}
  \caption{(Color online) (a) Current-voltage and (c) capacitance-voltage characteristics of \LSMO/(\SMO)$_x$/\NSTO \ junctions at room temperature for different interface compositions.  (b), (d) Histograms of the obtained SBH and $V_{bi}$ from (a) and (c) in steps of 0.04 V normalized by the total number of measurements. The curves show best Gaussian fits.}
  \label{fig:Fig2.eps}
\end{figure}
Figure \ref{fig:Fig2.eps}(a) shows typical \IV \ characteristics on a semi-logarithmic scale for samples with different \SMO \  coverage. Clear rectifying behavior was observed in all cases with forward biased current density systematically decreasing with increase in \SMO \ coverage. The barrier height obtained from the \IV \ characteristics ($\Phi_{SB}^{IV}$) was calculated based on thermoionic emission by fitting the forward biased region of the \IV \ characteristics. Here a Richardson constant of 156 AK$^{-2}$ was used ~\cite{Sroubek}. In order to obtain reliable statistics, 12 - 19 junctions were sampled for each composition. The obtained SBHs are summarized in a histogram shown in Fig. \ref{fig:Fig2.eps}(b) from which it is apparent that an increase in the SBH is observed as a function of \SMO \ coverage.  Considering the exponential dependence of the current on the SBH, the increase in the barrier height is consistent with the systematic decrease in the forward biased current shown in Fig. \ref{fig:Fig2.eps}(a).
\\\indent
The reverse biased junction capacitance characteristics at 1 kHz are presented in a $1/C^2$ - $V$ plot as shown in Fig. \ref{fig:Fig2.eps}(c). No frequency dependence was found for the capacitance from 20 Hz - 10 kHz, above which the junction \textit{RC} roll-off was observed. All samples showed a linear dependence on the applied voltage, from which the built-in potential ($V_{bi}$) was calculated, as  summarized in a histogram shown in Fig. \ref{fig:Fig2.eps}(d). The variance in $V_{bi}$ is smaller than that in $\Phi_{SB}^{IV}$, which is a reasonable consequence of the difference in measurement technique. The charge modulation at the edge of the depletion region far away from the interface in \CV \ tends to capture the spatial average of the barrier height, whereas in \IV\, the carriers surmount the interface barrier, making it more sensitive to the spatial distribution of the potential at the interface.
\begin{figure}[t]
  \begin{center}
    \includegraphics[clip]{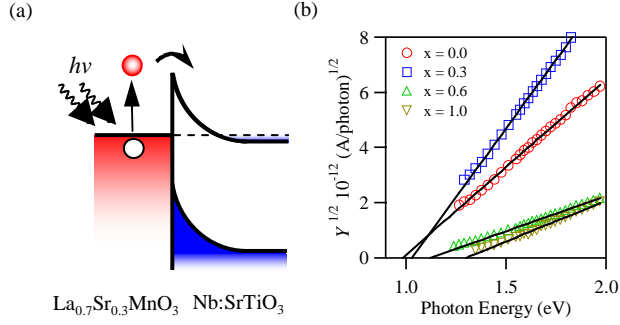}
  \end{center}
  \caption{(Color online) (a) Schematic illustration of internal photoemission (IPE). The electrons surmounting the barrier height are detected as photocurrent. (b) IPE spectra of \LSMO/(\SMO)$_x$/\NSTO \ junctions at room temperature. The square root of the photoyield is plotted against the photon energy.}
  \label{fig:Fig3.eps}
\end{figure}
\\\indent
Based on these results, IPE was measured directly through the \LSMO\ film (the Au film and electrode was mounted at the edge of the junction) for each composition as  shown in Fig. \ref{fig:Fig3.eps}. The details of the experimental configuration have been given previously ~\cite{Hikita}. The square root of the photoyield $\sqrt{Y}$, the photocurrent normalized by the incident photon count, is plotted against the incident photon energy. All samples exhibited a linear response of $\sqrt{Y}$, justifying the application of Fowler's equation to the emission process ~\cite{Fowler}, from which the barrier height $\Phi^{IPE}_{SB}$ is extrapolated.
\\\indent
The SBHs obtained from the three independent measurements are summarized in Fig. \ref{fig:Fig4.eps}. For \IV \ and \CV, the mean values obtained from the Gaussian fits to the histograms were used. Note that the barrier heights extracted from V$_{bi}$ determined by \CV \ measurements $\Phi^{CV}_{SB}$ have been corrected for the energy difference between the conduction band minimum and the Fermi level in the \NSTO\, as discussed in Ref. ~\cite{Hikita}, which is a small correction here ($\sim$ 8.1 mV). All measurements exhibit a systematic  increase in the SBH as a function of \SMO \ coverage at the interface. Although $\Phi^{IV}_{SB}$ commonly underestimates the SBH due to tunneling contributions or barrier inhomogeneities, the large discrepancy between $\Phi^{CV}_{SB}$ and $\Phi^{IPE}_{SB}$ is in contrast to the close correspondence of these measurements found for SrRuO$_3$/Nb:SrTiO$_3$ junctions ~\cite{Hikita}. A similar contrast between La$_{0.6}$Sr$_{0.4}$MnO$_3$/\NSTO\ and SrRuO$_3$/\NSTO\ interfaces was observed using photoemission spectroscopy ~\cite{Minohara}. For our data, the lack of a low-frequency dispersion to the capacitance indicates that we are not dominated by low-lying trap states. Nevertheless, the quantitative difference between $\Phi^{CV}_{SB}$ and $\Phi^{IPE}_{SB}$ indicates the presence of fluctuating dipoles at the interface. We turn now to consider the origin of these dipoles, and the systematic increase in the SBH.
\begin{figure}[t]
  \begin{center}
    \includegraphics[clip]{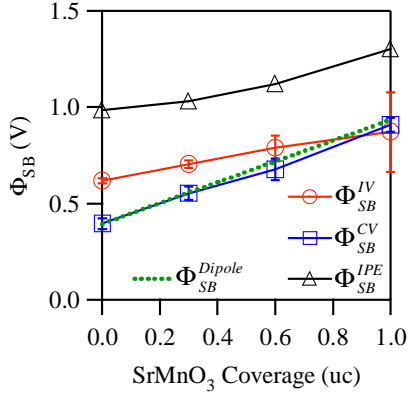}
  \end{center}
  \caption{(Color online) Summary of the obtained barrier heights from \IV\ (›, red), \CV\ ( , blue), and IPE (¢, black) measurements. The dotted line (green) indicates the variation of the screening dipole in a simple ionic model (see text for details).}
  \label{fig:Fig4.eps}
\end{figure}
\\\indent
First we consider established semiconductor models for Schottky barrier formation. In the simplest Schottky-Mott model of a metal-semiconductor junction, the SBH is purely determined by the difference in the work function of the metal ($\Phi_M$) and the electron affinity ($\chi$) of the semiconductor ~\cite{Sze}, and hence cannot capture any termination dependence of the SBH. The classical Fermi level pinning mechanism based on the surface states of semiconductors proposed by Bardeen ~\cite{Bardeen} does not capture variations in metal screening discussed below. Recently, a bond polarization model has been developed ~\cite{Tung}, which incorporates both bulk and interface contributions. The interface specific properties are incorporated by electric dipoles generated by the abrupt break in the periodicity of the crystal potential. For an interface between a metal and an \textit{n}-type semiconductor, the SBH is expressed as ~\cite{Tung},
\begin{equation}
\Phi_{SB} =\gamma_B\left(\Phi_M-\chi\right) + \left(1-\gamma_B\right)\frac{E_g}{2}, 
\end{equation}
where
\begin{equation}
\gamma_B = 1 - \frac{q^2N_Bd_{MS}}{\epsilon_{it}\left(E_g+\kappa\right)}.
\end{equation}
Here $E_g$ is the semiconductor band-gap, $N_B$ the number of interface metal-semiconductor bonds (dipoles), $d_{MS}$ the metal-semiconductor bonding distance, $\epsilon_{it}$ the dielectric constant at the interface, $\kappa$ the specific Coulomb interaction between the neighboring atoms at the interface, and $q$ the electronic charge. Given that the possible range of $d_{MS}$ is too small to account for the shifts we have observed, $N_B$ is the only parameter linearly varying the SBH, and it is reasonable to conclude that the dipole density at the interface, or the strength of the interface dipole, decreased with \SMO \ coverage. The limitation of the bond polarization model for our purposes is that it was established for application to covalent semiconductor interfaces, in which the concept of a number of "chemical bonds" at the interface is justified. However, for more ionic semiconductors, such as the case here, the concept of chemical bonds becomes ambiguous because the cohesion of the lattice is dominated by the Madelung energy in the whole crystal rather than local atomic bonds. We discuss below one possible picture of the interface dipole by considering the polarity mismatch at the interface.
\begin{figure}[t]
  \begin{center}
    \includegraphics[clip]{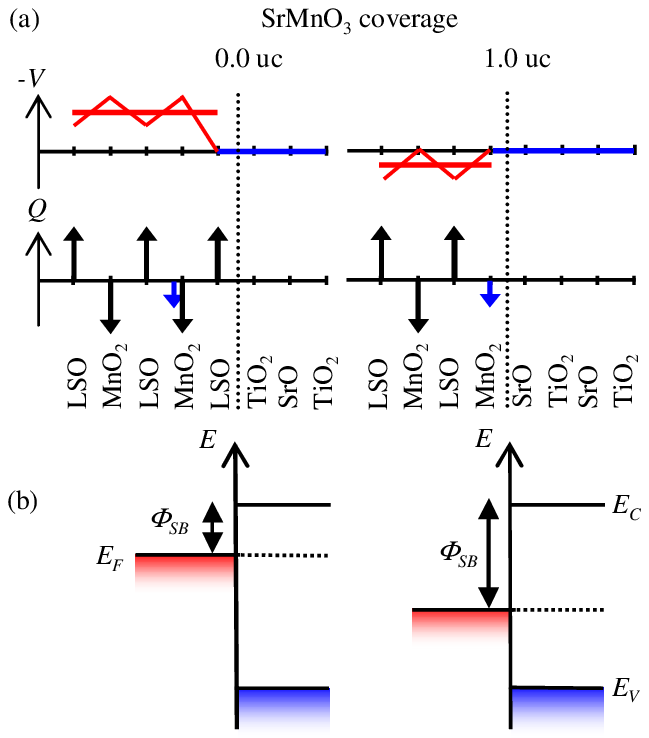}
  \end{center}
  \caption{(Color online) (a) A schematic diagram of the \LSMO/\NSTO \ interface charge sheet density and the electrostatic potential for 0.0 uc  (left) and 1.0 uc (right) of \SMO\ coverage. The small arrows in \LSMO \ represent the compensation charges  induced to screen the interface. The relative electrostatic potential across the interface varies depending on the interface termination, consequently changing the band alignment (b) at the interface (see text for details).}
  \label{fig:Fig5.eps}
\end{figure}
\\\indent
Because the present interface systematically changes from MnO$_2$/\textit{\LS}/TiO$_2$ to MnO$_2$/\textit{SrO}/TiO$_2$, the sheet charge density shifts from -0.7\textit{q} / +0.7\textit{q} / 0\textit{q} to -0.7\textit{q} / 0\textit{q} / 0\textit{q}, assuming a fully ionic charge assignment using the nominal bulk valence for each grown layer, creating a polar discontinuity at the interface. In order to avoid a diverging electrostatic potential arising from the interface, -/+0.35\textit{q} extra charge is required at the two interfaces, respectively.  Whereas previous considerations of this effect between two insulators were discussed in terms of electronic reconstructions ~\cite{SOkamoto1, ZSPopvic, Ohtomo, WCLee, RPentcheva, ARuegg, Smadici} , here the interface between a metal and a semiconductor is better framed in terms of metallic screening by the \LSMO\ --- the \NSTO\ side of the interface being fully depleted. \\\indent
To estimate the length scale for screening, the Thomas-Fermi screening length is $\sim$ 0.31 nm, using a bulk carrier density of 5.1 $\times$ 10$^{21}$ cm$^{-3}$, a dielectric constant $\epsilon$ of 30 ~\cite{Shannon}, and an electron effective mass of 2.5 ~\cite{Okuda}. This length scale, less than a unit cell, corresponds to changing the valence of Mn at the first interface layer in the simplest ionic assignment. Thus, as depicted in Fig. \ref{fig:Fig5.eps}, the first MnO$_2$ layer of \LSMO\ will have extra screening charge. Even after this charge compensation, a finite electrostatic potential remains inside \LSMO\ relative to \NSTO, giving an interface dipole which linearly varies with the interface termination.
\\\indent
The variation in the band-offset induced by the difference in the termination at the interface can be estimated using the charge assignment shown in Fig. \ref{fig:Fig5.eps}. Using the previous values used for the Thomas-Fermi estimate, the evolution of the SBH arising from this ionic dipole is given in Fig. \ref{fig:Fig4.eps}, referenced to the Schottky-Mott relation ~\cite{Minohara}. The electrostatic potential difference between the two end-member interfaces is 0.54 V. This value, as well as the linearly increasing SBH with varying interface termination, are in reasonable agreement with the experimentally determined trends.
\\\indent
In summary, we have presented experiments finding a systematic increase in the Schottky barrier height in \LSMO/\NSTO\ (001) heterojunctions as the \NSTO\ semiconductor was varied from TiO$_2$ termination to SrO termination, and a simple model for interface dipole formation which captures this trend.  It should be noted that the ionic limit discussed here is just as oversimplified as the covalent limit used in the bond polarization model; the real system is intermediate between these two extremes. In addition to hybridization effects, a more realistic estimate of the interface dipole requires better understanding of the relevant $\epsilon$ on these very short length scales. \textit{Ab initio} calculations such as recently performed for ultrathin perovskite superlattices should give more quantitative insight ~\cite{Spaldin, Hamann}. Nevertheless, this basic framework for interface dipole formation is quite general, and should assist in the design of oxide heterostructures and control of their band alignments.
\\\indent
This work was supported by the TEPCO Research Foundation and a Grant-in-Aid for Scientific Research on Priority Areas.

\begin{thebibliography}{26}
\expandafter\ifx\csname natexlab\endcsname\relax\def\natexlab#1{#1}\fi
\expandafter\ifx\csname bibnamefont\endcsname\relax
  \def\bibnamefont#1{#1}\fi
\expandafter\ifx\csname bibfnamefont\endcsname\relax
  \def\bibfnamefont#1{#1}\fi
\expandafter\ifx\csname citenamefont\endcsname\relax
  \def\citenamefont#1{#1}\fi
\expandafter\ifx\csname url\endcsname\relax
  \def\url#1{\texttt{#1}}\fi
\expandafter\ifx\csname urlprefix\endcsname\relax\def\urlprefix{URL }\fi
\providecommand{\bibinfo}[2]{#2}
\providecommand{\eprint}[2][]{\url{#2}}

\bibitem[{\citenamefont{Okamoto and Millis}(2004)}]{SOkamoto1}
\bibinfo{author}{\bibfnamefont{S.}~\bibnamefont{Okamoto}} \bibnamefont{and}
  \bibinfo{author}{\bibfnamefont{A.~J.} \bibnamefont{Millis}},
  \bibinfo{journal}{Nature (London)} \textbf{\bibinfo{volume}{428}},
  \bibinfo{pages}{630} (\bibinfo{year}{2004}).

\bibitem[{\citenamefont{Ohtomo and Hwang}(2004)}]{Ohtomo}
\bibinfo{author}{\bibfnamefont{A.}~\bibnamefont{Ohtomo}} \bibnamefont{and}
  \bibinfo{author}{\bibfnamefont{H.~Y.} \bibnamefont{Hwang}},
  \bibinfo{journal}{Nature} \textbf{\bibinfo{volume}{427}},
  \bibinfo{pages}{423} (\bibinfo{year}{2004}).

\bibitem[{\citenamefont{Popovic and Satpathy}(2005)}]{ZSPopvic}
\bibinfo{author}{\bibfnamefont{Z.~S.} \bibnamefont{Popovic}} \bibnamefont{and}
  \bibinfo{author}{\bibfnamefont{S.}~\bibnamefont{Satpathy}},
  \bibinfo{journal}{Phys. Rev. Lett.} \textbf{\bibinfo{volume}{94}},
  \bibinfo{pages}{176805} (\bibinfo{year}{2005}).

\bibitem[{\citenamefont{Lee and MacDonald}(2006)}]{WCLee}
\bibinfo{author}{\bibfnamefont{W.~C.} \bibnamefont{Lee}} \bibnamefont{and}
  \bibinfo{author}{\bibfnamefont{A.~H.} \bibnamefont{MacDonald}},
  \bibinfo{journal}{Phys. Rev. B} \textbf{\bibinfo{volume}{74}},
  \bibinfo{pages}{075106} (\bibinfo{year}{2006}).

\bibitem[{\citenamefont{Pentcheva and Pickett}(2007)}]{RPentcheva}
\bibinfo{author}{\bibfnamefont{R.}~\bibnamefont{Pentcheva}} \bibnamefont{and}
  \bibinfo{author}{\bibfnamefont{W.~E.} \bibnamefont{Pickett}},
  \bibinfo{journal}{Phys. Rev. Lett.} \textbf{\bibinfo{volume}{99}},
  \bibinfo{pages}{016802} (\bibinfo{year}{2007}).

\bibitem[{\citenamefont{Reyren et~al.}(2007)\citenamefont{Reyren, Thiel,
  Caviglia, Kourkoutis, Hammerl, Richter, Schneider, Kopp, Ruetschi, Jaccard
  et~al.}}]{NReynen}
\bibinfo{author}{\bibfnamefont{N.}~\bibnamefont{Reyren}},
  \bibinfo{author}{\bibfnamefont{S.}~\bibnamefont{Thiel}},
  \bibinfo{author}{\bibfnamefont{A.~D.} \bibnamefont{Caviglia}},
  \bibinfo{author}{\bibfnamefont{L.~F.} \bibnamefont{Kourkoutis}},
  \bibinfo{author}{\bibfnamefont{G.}~\bibnamefont{Hammerl}},
  \bibinfo{author}{\bibfnamefont{C.}~\bibnamefont{Richter}},
  \bibinfo{author}{\bibfnamefont{C.~W.} \bibnamefont{Schneider}},
  \bibinfo{author}{\bibfnamefont{T.}~\bibnamefont{Kopp}},
  \bibinfo{author}{\bibfnamefont{A.-S.} \bibnamefont{Ruetschi}},
  \bibinfo{author}{\bibfnamefont{D.}~\bibnamefont{Jaccard}},
  \bibnamefont{et~al.}, \bibinfo{journal}{Science}
  \textbf{\bibinfo{volume}{317}}, \bibinfo{pages}{1196} (\bibinfo{year}{2007}).

\bibitem[{\citenamefont{Ruegg et~al.}(2007)\citenamefont{Ruegg, Pilgram, and
  Sigrist}}]{ARuegg}
\bibinfo{author}{\bibfnamefont{A.}~\bibnamefont{Ruegg}},
  \bibinfo{author}{\bibfnamefont{S.}~\bibnamefont{Pilgram}}, \bibnamefont{and}
  \bibinfo{author}{\bibfnamefont{M.}~\bibnamefont{Sigrist}},
  \bibinfo{journal}{Phys. Rev. B} \textbf{\bibinfo{volume}{75}},
  \bibinfo{pages}{195117} (\bibinfo{year}{2007}).

\bibitem[{\citenamefont{Smadici et~al.}(2007)\citenamefont{Smadici, Abbamonte,
  Bhattacharya, Zhai, Jiang, Rusydi, Eckstein, Bader, and Zuo}}]{Smadici}
\bibinfo{author}{\bibfnamefont{S.}~\bibnamefont{Smadici}},
  \bibinfo{author}{\bibfnamefont{P.}~\bibnamefont{Abbamonte}},
  \bibinfo{author}{\bibfnamefont{A.}~\bibnamefont{Bhattacharya}},
  \bibinfo{author}{\bibfnamefont{X.}~\bibnamefont{Zhai}},
  \bibinfo{author}{\bibfnamefont{B.}~\bibnamefont{Jiang}},
  \bibinfo{author}{\bibfnamefont{A.}~\bibnamefont{Rusydi}},
  \bibinfo{author}{\bibfnamefont{J.~N.} \bibnamefont{Eckstein}},
  \bibinfo{author}{\bibfnamefont{S.~D.} \bibnamefont{Bader}}, \bibnamefont{and}
  \bibinfo{author}{\bibfnamefont{J.-M.} \bibnamefont{Zuo}},
  \bibinfo{journal}{Phys. Rev. Lett.} \textbf{\bibinfo{volume}{99}},
  \bibinfo{pages}{196404} (\bibinfo{year}{2007}).

\bibitem[{\citenamefont{Chaloupka and Khaliullin}(2008)}]{JChaloupka}
\bibinfo{author}{\bibfnamefont{J.}~\bibnamefont{Chaloupka}} \bibnamefont{and}
  \bibinfo{author}{\bibfnamefont{G.}~\bibnamefont{Khaliullin}},
  \bibinfo{journal}{Phys. Rev. Lett.} \textbf{\bibinfo{volume}{100}},
  \bibinfo{pages}{016404} (\bibinfo{year}{2008}).

\bibitem[{\citenamefont{Yunoki et~al.}(2007)\citenamefont{Yunoki, Moreo,
  Dagotto, Okamoto, Kancharla, and Fujimori}}]{Okamoto}
\bibinfo{author}{\bibfnamefont{S.}~\bibnamefont{Yunoki}},
  \bibinfo{author}{\bibfnamefont{A.}~\bibnamefont{Moreo}},
  \bibinfo{author}{\bibfnamefont{E.}~\bibnamefont{Dagotto}},
  \bibinfo{author}{\bibfnamefont{S.}~\bibnamefont{Okamoto}},
  \bibinfo{author}{\bibfnamefont{S.~S.} \bibnamefont{Kancharla}},
  \bibnamefont{and} \bibinfo{author}{\bibfnamefont{A.}~\bibnamefont{Fujimori}},
  \bibinfo{journal}{Phys. Rev. B} \textbf{\bibinfo{volume}{76}},
  \bibinfo{pages}{064532} (\bibinfo{year}{2007}).

\bibitem[{\citenamefont{Sun et~al.}(1996)\citenamefont{Sun, Gallagher,
  Duncombe, Krusin-Elbaum, Altman, Gupta, Lu, Gong, and Xiao}}]{JZSun}
\bibinfo{author}{\bibfnamefont{J.~Z.} \bibnamefont{Sun}},
  \bibinfo{author}{\bibfnamefont{W.~J.} \bibnamefont{Gallagher}},
  \bibinfo{author}{\bibfnamefont{P.~R.} \bibnamefont{Duncombe}},
  \bibinfo{author}{\bibfnamefont{L.}~\bibnamefont{Krusin-Elbaum}},
  \bibinfo{author}{\bibfnamefont{R.~A.} \bibnamefont{Altman}},
  \bibinfo{author}{\bibfnamefont{A.}~\bibnamefont{Gupta}},
  \bibinfo{author}{\bibfnamefont{Y.}~\bibnamefont{Lu}},
  \bibinfo{author}{\bibfnamefont{G.~Q.} \bibnamefont{Gong}}, \bibnamefont{and}
  \bibinfo{author}{\bibfnamefont{G.}~\bibnamefont{Xiao}},
  \bibinfo{journal}{Appl. Phys. Lett.} \textbf{\bibinfo{volume}{69}},
  \bibinfo{pages}{3266} (\bibinfo{year}{1996}).

\bibitem[{\citenamefont{Nakagawa et~al.}(2005)\citenamefont{Nakagawa, Asai,
  Mukunoki, Susaki, and Hwang}}]{Nakagawa}
\bibinfo{author}{\bibfnamefont{N.}~\bibnamefont{Nakagawa}},
  \bibinfo{author}{\bibfnamefont{M.}~\bibnamefont{Asai}},
  \bibinfo{author}{\bibfnamefont{Y.}~\bibnamefont{Mukunoki}},
  \bibinfo{author}{\bibfnamefont{T.}~\bibnamefont{Susaki}}, \bibnamefont{and}
  \bibinfo{author}{\bibfnamefont{H.~Y.} \bibnamefont{Hwang}},
  \bibinfo{journal}{Appl. Phys. Lett.} \textbf{\bibinfo{volume}{86}},
  \bibinfo{pages}{082504} (\bibinfo{year}{2005}).

\bibitem[{\citenamefont{Katsu et~al.}(2000)\citenamefont{Katsu, Tanaka, and
  Kawai}}]{Katsu}
\bibinfo{author}{\bibfnamefont{H.}~\bibnamefont{Katsu}},
  \bibinfo{author}{\bibfnamefont{H.}~\bibnamefont{Tanaka}}, \bibnamefont{and}
  \bibinfo{author}{\bibfnamefont{T.}~\bibnamefont{Kawai}},
  \bibinfo{journal}{Appl. Phys. Lett.} \textbf{\bibinfo{volume}{76}},
  \bibinfo{pages}{3245} (\bibinfo{year}{2000}).

\bibitem[{\citenamefont{Hotta et~al.}(2007)\citenamefont{Hotta, Susaki, and
  Hwang}}]{Hotta}
\bibinfo{author}{\bibfnamefont{Y.}~\bibnamefont{Hotta}},
  \bibinfo{author}{\bibfnamefont{T.}~\bibnamefont{Susaki}}, \bibnamefont{and}
  \bibinfo{author}{\bibfnamefont{H.~Y.} \bibnamefont{Hwang}},
  \bibinfo{journal}{Phys. Rev. Lett.} \textbf{\bibinfo{volume}{99}},
  \bibinfo{pages}{236805} (\bibinfo{year}{2007}).

\bibitem[{\citenamefont{Sze}(1981)}]{Sze}
\bibinfo{author}{\bibfnamefont{S.~M.} \bibnamefont{Sze}},
  \emph{\bibinfo{title}{Physics of Semiconductor Devices, 2nd ed.}}
  (\bibinfo{publisher}{Wiley, New York}, \bibinfo{year}{1981}).

\bibitem[{\citenamefont{Bardeen}(1947)}]{Bardeen}
\bibinfo{author}{\bibfnamefont{J.}~\bibnamefont{Bardeen}},
  \bibinfo{journal}{Phys. Rev.} \textbf{\bibinfo{volume}{71}},
  \bibinfo{pages}{717} (\bibinfo{year}{1947}).

\bibitem[{\citenamefont{Song et~al.}(2008)\citenamefont{Song, Susaki, and
  Hwang}}]{Song}
\bibinfo{author}{\bibfnamefont{J.~H.} \bibnamefont{Song}},
  \bibinfo{author}{\bibfnamefont{T.}~\bibnamefont{Susaki}}, \bibnamefont{and}
  \bibinfo{author}{\bibfnamefont{H.~Y.} \bibnamefont{Hwang}},
  \bibinfo{journal}{Adv. Mater.} \textbf{\bibinfo{volume}{20}},
  \bibinfo{pages}{2528} (\bibinfo{year}{2008}).

\bibitem[{\citenamefont{Sroubek}(1970)}]{Sroubek}
\bibinfo{author}{\bibfnamefont{Z.}~\bibnamefont{Sroubek}},
  \bibinfo{journal}{Phys. Rev. B} \textbf{\bibinfo{volume}{2}},
  \bibinfo{pages}{3170} (\bibinfo{year}{1970}).

\bibitem[{\citenamefont{Hikita et~al.}(2007)\citenamefont{Hikita, Kozuka,
  Susaki, Takagi, and Hwang}}]{Hikita}
\bibinfo{author}{\bibfnamefont{Y.}~\bibnamefont{Hikita}},
  \bibinfo{author}{\bibfnamefont{Y.}~\bibnamefont{Kozuka}},
  \bibinfo{author}{\bibfnamefont{T.}~\bibnamefont{Susaki}},
  \bibinfo{author}{\bibfnamefont{H.}~\bibnamefont{Takagi}}, \bibnamefont{and}
  \bibinfo{author}{\bibfnamefont{H.~Y.} \bibnamefont{Hwang}},
  \bibinfo{journal}{Appl. Phys. Lett.} \textbf{\bibinfo{volume}{90}},
  \bibinfo{pages}{143507} (\bibinfo{year}{2007}).

\bibitem[{\citenamefont{Fowler}(1931)}]{Fowler}
\bibinfo{author}{\bibfnamefont{R.}~\bibnamefont{Fowler}},
  \bibinfo{journal}{Phys. Rev.} \textbf{\bibinfo{volume}{38}},
  \bibinfo{pages}{45} (\bibinfo{year}{1931}).

\bibitem[{\citenamefont{Minohara et~al.}(2007)\citenamefont{Minohara, Ohkubo,
  Kumigashira, and Oshima}}]{Minohara}
\bibinfo{author}{\bibfnamefont{M.}~\bibnamefont{Minohara}},
  \bibinfo{author}{\bibfnamefont{I.}~\bibnamefont{Ohkubo}},
  \bibinfo{author}{\bibfnamefont{H.}~\bibnamefont{Kumigashira}},
  \bibnamefont{and} \bibinfo{author}{\bibfnamefont{M.}~\bibnamefont{Oshima}},
  \bibinfo{journal}{Appl. Phys. Lett.} \textbf{\bibinfo{volume}{90}},
  \bibinfo{pages}{132123} (\bibinfo{year}{2007}).

\bibitem[{\citenamefont{Tung}(2001)}]{Tung}
\bibinfo{author}{\bibfnamefont{R.}~\bibnamefont{Tung}},
  \bibinfo{journal}{Mater. Sci. Eng. R} \textbf{\bibinfo{volume}{35}},
  \bibinfo{pages}{1} (\bibinfo{year}{2001}).

\bibitem[{\citenamefont{Shannon}(1993)}]{Shannon}
\bibinfo{author}{\bibfnamefont{R.~D.} \bibnamefont{Shannon}},
  \bibinfo{journal}{J. Appl. Phys.} \textbf{\bibinfo{volume}{73}},
  \bibinfo{pages}{348} (\bibinfo{year}{1993}).

\bibitem[{\citenamefont{Okuda et~al.}(1998)\citenamefont{Okuda, Asamitsu,
  Tomioka, Kimura, Taguchi, and Tokura}}]{Okuda}
\bibinfo{author}{\bibfnamefont{T.}~\bibnamefont{Okuda}},
  \bibinfo{author}{\bibfnamefont{A.}~\bibnamefont{Asamitsu}},
  \bibinfo{author}{\bibfnamefont{Y.}~\bibnamefont{Tomioka}},
  \bibinfo{author}{\bibfnamefont{T.}~\bibnamefont{Kimura}},
  \bibinfo{author}{\bibfnamefont{Y.}~\bibnamefont{Taguchi}}, \bibnamefont{and}
  \bibinfo{author}{\bibfnamefont{Y.}~\bibnamefont{Tokura}},
  \bibinfo{journal}{Phys. Rev. Lett.} \textbf{\bibinfo{volume}{81}},
  \bibinfo{pages}{3203} (\bibinfo{year}{1998}).

\bibitem[{\citenamefont{Stengel and Spaldin}(2006)}]{Spaldin}
\bibinfo{author}{\bibfnamefont{M.}~\bibnamefont{Stengel}} \bibnamefont{and}
  \bibinfo{author}{\bibfnamefont{N.~A.} \bibnamefont{Spaldin}},
  \bibinfo{journal}{Nature} \textbf{\bibinfo{volume}{443}},
  \bibinfo{pages}{679} (\bibinfo{year}{2006}).

\bibitem[{\citenamefont{Hamann et~al.}(2006)\citenamefont{Hamann, Muller, and
  Hwang}}]{Hamann}
\bibinfo{author}{\bibfnamefont{D.~R.} \bibnamefont{Hamann}},
  \bibinfo{author}{\bibfnamefont{D.~A.} \bibnamefont{Muller}},
  \bibnamefont{and} \bibinfo{author}{\bibfnamefont{H.~Y.} \bibnamefont{Hwang}},
  \bibinfo{journal}{Phys. Rev. B} \textbf{\bibinfo{volume}{73}},
  \bibinfo{pages}{195403} (\bibinfo{year}{2006}).

\end{thebibliography}

\end{document}